# Higher superconducting transition temperature by breaking the universal pressure relation

Liangzi Deng[1], Yongping Zheng[2], Zheng Wu[1], Shuyuan Huyan[1], Hung-Cheng Wu[1], Yifan Nie[2], Kyeongjae Cho[2], and Ching-Wu Chu[1,3*]

[1]Department of Physics and Texas Center for Superconductivity, University of Houston, Houston TX 77204
[2]Department of Materials Science & Engineering, University of Texas at Dallas, Richardson TX 75080
[3]Lawrence Berkeley National Laboratory, 1 Cyclotron Road, Berkeley CA 94720
*email: cwchu@uh.edu

**Abstract**: By investigating the bulk superconducting state via dc magnetization measurements, we have discovered a common resurgence of the superconductive transition temperatures ($T_c$s) of the monolayer $Bi_2Sr_2CuO_{6+\delta}$ (Bi2201) and bilayer $Bi_2Sr_2CaCu_2O_{8+\delta}$ (Bi2212) to beyond the maximum $T_c$s ($T_{c-max}$s) predicted by the universal relation between $T_c$ and doping ($p$) or pressure (P) at higher pressures. The $T_c$ of under-doped Bi2201 initially increases from 9.6 K at ambient to a peak at ~ 23 K at ~ 26 GPa and then drops as expected from the universal $T_c$-P relation. However, at pressures above ~ 40 GPa, $T_c$ rises rapidly without any sign of saturation up to ~ 30 K at ~ 51 GPa. Similarly, the $T_c$ for the slightly overdoped Bi2212 increases after passing a broad valley between 20-36 GPa and reaches ~ 90 K without any sign of saturation at ~ 56 GPa. We have therefore attributed this $T_c$-resurgence to a possible pressure-induced electronic transition in the cuprate compounds due to a charge transfer between the Cu $3d_{x^2-y^2}$ and the O $2p$ bands projected from a hybrid bonding state, leading to an increase of the density of states at the Fermi level, in agreement with our density functional theory calculations. Similar $T_c$-P behavior has also been reported in the trilayer $Br_2Sr_2Ca_2Cu_3O_{10+\delta}$ (Bi2223). These observations suggest that higher $T_c$s than those previously reported for the layered cuprate high temperature superconductors can be achieved by breaking away from the universal $T_c$-P relation through the application of higher pressures.

**Significance Statement**: Achieving higher transition temperature ($T_c$) is a primary goal in superconductivity research. $T_c$ and doping have been found to have a dome-like universal relation where the peak position is the maximum $T_{c-max}$, which is consistent with previous experimental results in the lower pressure range. By using our newly developed ultra-sensitive magnetization measurement technique under high pressure, we discovered a universal resurgence of $T_c$ passing the peak predicted by the general $T_c$-$p$ (doping) or -P (pressure) relation for cuprate high temperature superconductor and attribute the resurgence to a pressure-induced electronic transition, which is supported qualitatively by our density functional theory calculations. This offers a new way to raise the $T_c$ of the layered cuprate high temperature superconductors to a new height.

# Introduction

Raising the transition temperature ($T_c$) of superconductors has been a primary driving force behind the sustained effort in superconductivity research ever since its discovery due to the scientific challenges and technological promise that superconductivity at higher temperature can offer. This has been particularly true since the discovery of high temperature superconductivity (HTS) in the late 1980s (1, 2). The most promising stable HTS systems for higher $T_c$ appear to be the layered cuprates (3), Fe-pnictides (4), and Fe-chalcogenides (4). To date, all known stable HTSs with a $T_c$ above 77 K, the liquid nitrogen boiling point, at ambient or high pressure, are perovskite-like layered cuprates. For example, $HgBa_2Ca_2Cu_3O_{8+\delta}$ exhibits the current record $T_c$s of 134 K (5) at ambient and 164 K above 30 GPa (6, 7), respectively. Superconductivity determined resistively has been reported at up to 109 K in unit-cell FeSe/STO film at ambient (8), but not yet confirmed or reproduced; and at up to 203 K under ~ 155 GPa (9) and ~ 260 K under ~ 190 GPa (10, 11) in $H_3S$ and $LaH_{10+x}$, respectively, but these compounds are unstable.

Superconductivity can be induced in the stable cuprates from their Mott insulating parents by chemical, pressure, photon (12), or electric-field doping (13). It has been found that, independent of the type of doping, the normalized transition temperature ($T_c/T_{c-max}$) of cuprates varies parabolically with the hole concentration ($p$) in a universal manner, *i.e.* ($T_c/T_{c-max}$) = [1- 82.6($p - p_o$)$^2$] (14), where $T_{c-max}$ is the maximum $T_c$ of a specific compound series when optimally doped with $p = p_o$ ~ 0.16 hole/$CuO_2$-layer. This is in general agreement with the rigid-band model when only small perturbations occur to the electron band structures in their equilibrium state through doping, and doping mainly shifts the Fermi energy, as has been demonstrated, e.g., in the $HgBa_2Ca_{n-1}Cu_nO_{2n+2+\delta}$ compound family for n =1, 2, and 3, i.e. for 1, 2, and 3 $CuO_2$-layers per unit cell. The observation of this universal $T_c/T_{c-max}$-$p$ (or -P) behavior suggests that the usual doping (chemical $p$ or pressure P) of cuprates in their equilibrium state cannot lead to a $T_c$ higher than $T_{c-max}$. To achieve a $T_c$ higher than $T_{c-max}$ in cuprates, one may have to modify their electronic structures, such as by inducing a Fermi surface topology change (15, 16) via very high pressures, i.e. P >> ~ 20 GPa, the pressure that most previous experiments on cuprates have reached. In other words, we have to break away from the commonly accepted universal $T_c$-$p$ (or $T_c/T_{c-max}$-$p$ or $T_c/T_{c-max}$-P) relation.

To explore such a possibility, we have investigated the high-pressure effects on the $T_c$s of members of the homologous series of $Bi_2Sr_2Ca_{n-1}Cu_nO_{2n+4+\delta}$ (BSCCO) (Fig. 1d) with n = 1 and 2, where n is the number of $CuO_2$ layers per cell, magnetically up to ~ 56 GPa. Indeed, we have observed the resurgence of their $T_c$s at higher pressures after reaching the $T_{c-max}$ predicted by the universal quadratic ($T_c/T_{c-max}$-$p$) relation. Similar $T_c$-resurgence under high pressure has also been reported in the optimally doped Bi2223 with n = 3 by Chen et al. (17), although the authors attributed the $T_c$-rise to the possible pressure-induced competition between the pairing and phase ordering among the three different $CuO_2$ planes specific to Bi2223. Our experiments show that the resurgence of $T_c$ beyond the $T_{c-max}$ clearly is n-independent and appears to be common to layered cuprates, probably due to a pressure-induced electronic transition that gives rise to the increase of the electron density of states at the Fermi level, consistent with our density functional theory (DFT) calculations.

Pressure has played a crucial role in the development of HTS due to its simplicity in varying the basic parameter of solids, the inter-atomic distance, without introducing complications associated with altering the chemistry of the compound. The close relationship between pressure and doping has been demonstrated experimentally and explained theoretically (18, 19). The universal $T_c$-$p$ relation can therefore be replaced by the universal $T_c$-P relation. For instance, it was the anomalously large positive pressure effect on the $T_c$ of the first cuprate superconductor, $(La,Ba)_2CuO_4$ (20, 21), that prompted the substitution of the smaller ion $Y^{3+}$ for $La^{3+}$ to generate lattice pressure, leading to the discovery of the first liquid-nitrogen HTS, $YBa_2Cu_3O_{7+\delta}$ (2). It was also the high pressure of ~ 31 GPa that helped achieve the current record-high $T_c$ in $HgBa_2Ca_2Cu_3O_{8+\delta}$ (6, 7). Pressure on a compound can change its carrier concentration and shift its Fermi level, and can sometimes even alter its Fermi surface topology, leading to a Lifshitz transition (15, 16). It has been demonstrated that $dlnT_c/dP$ is positive for under-doped HTSs, negative for over-doped HTSs, and ~ 0 for optimally doped HTSs (22), in agreement with the universal $T_c(p)$, at least at low pressure, e.g. < 20 GPa. This is because, within the framework of the rigid-band model, $dT_c/dP$ can be written as $dT_c/dP = (dT_c/dp)(dp/dP)$. By assuming a positive $dp/dP$, the sign change of $dT_c/dP$ with $p$ is determined by the sign of the slope of $T_c(p)$ and can thus be understood.

## Results

Recently, we have developed an ultra-sensitive high-pressure miniature diamond-anvil-cell (mini-DAC) technique (Fig. 1a) that enables us to investigate the bulk superconducting state of a solid with ≤ 100 μm diagonal by directly measuring the DC magnetization under pressures up to ~ 60 GPa. We used this technique to investigate the bulk superconducting states of the under-doped Bi2201 and the nearly optimally doped Bi2212. The pressure dependences of $T_c$ are shown in Figs. 2 and 3. The $T_c$s were determined by DC magnetization measurements. As shown in Fig. 2, the $T_c$ of the under-doped Bi2201 initially increases rapidly from 9.6 K at ambient to a peak at ~ 23 K ($T_{c\text{-max}}$) at ~ 26 GPa and then drops to ~ 15 K at ~ 40 GPa, as expected for an under-doped cuprate HTS in accordance with the universal $T_c$-P relation. However, $T_c$ at pressures above 40 GPa rises rapidly without any sign of saturation up to 30 K at 51 GPa, the highest pressure applied, at a rate of ~ +1.5 K/GPa, in contrast to the continued drop with pressure predicted by the universal $T_c$-P relation. It is evident that the pressure effect observed is rather reversible. Fig. 3 shows the $T_c$-behavior under pressures reversibly for two single-crystalline samples of Bi2212. The sample is slightly overdoped as evidenced by its slightly lower $T_c$. The $T_c$-P behavior is qualitatively similar to that of Bi2201 at high pressures after reaching its $T_{c\text{-max}}$. Under pressure, $T_c$ decreases rapidly from ~ 90 K at ambient to ~ 71 K at ~20 GPa, as expected for an over-doped cuprate HTS based on the universal $T_c$-P relation; continues to decrease but at a slower rate above ~ 20 GPa; passes a broad minimum of ~ 69 K at ~ 30 GPa; and finally rises rapidly after ~ 40 GPa to ~ 90 K at ~ 56 GPa, the highest pressure applied, at a rate of ~ +1.4 K/GPa. As shown in the same figure, the low-pressure effect on the $T_c$ of our samples does not result in the peak previously reported by Chen et al. (23). We attribute the difference to the different doping levels of the samples studied; i.e., our samples are slightly over-doped, while the sample of Chen et al. is slightly under-doped, which may require a higher pressure to induce a resurgence of $T_c$. Our observations show that while the $T_c$ of Bi2201 and Bi2212 under low pressures (e.g. < 20 GPa) behaves in accordance with the universal $T_c$-P relation, a higher $T_c$ emerges at higher pressures, possibly associated with a subtle change of the electron energy structures, characteristic of the $CuO_2$-layers in cuprates. In fact, similar $T_c$-P behavior has also been reported in Bi2223 (17) with the triple $CuO_2$-layers per unit cell, i.e., under pressure its $T_c$ increases from 108 K at ambient, peaks at 123 K ($T_{c\text{-max}}$) at 12 GPa, decreases to 118 K at 23 GPa and finally increases to 136 K at 36 GPa at a rate of ~ +1.4 K/GPa. The final $T_c$-increase at high pressures was ascribed (17) by Chen et al. to a pressure-driven transition of the inner $CuO_2$-layer from the competing order to a superconducting state, thus specific to the triple $CuO_2$-layer cuprates. However, we attribute the $T_c$-rise at higher pressure after reaching its peak to an electronic transition generic to the $CuO_2$-layers in cuprates under very high pressures, independent of the number of $CuO_2$-layers per unit cell, as summarized in Table 1. It is interesting to note that following the $T_c$-resurgence for Bi2201, Bi2212, and Bi2223, their $dT_c/dP$ values are large and rather similar, but the sign of the $dT_c/dP$ at low pressures depends on the compound's initial doping state. The observations therefore suggest that higher $T_c$s than those previously reported in layered cuprate HTSs can be achieved by breaking away from the universal $T_c$-P relation via inducing a transition in the electronic spectrum of the $CuO_2$-layer by the application of higher pressures.

Numerous structural studies on BSCCO have been made under pressures up to 50 GPa at room temperature (17, 24-26). None have displayed any pressure-induced structure transition, although charge redistribution within the cell under pressure has been suggested from the Raman studies (17). The absence of any pressure-induced structure transition is consistent with our proposed electronic transition, such as the Lifshitz transition (15, 16), which may not be accompanied by a structural change. However, in spite of the absence of a structural transition, anomalies appear in the *c*-lattice parameter of Bi2212 at ~ 30 GPa (24), coinciding with the broad U-shaped $T_c$-valley we have observed, as shown in the Fig. 3 inset.

## Discussion

In searching for the microscopic origin of the unusual pressure-enhancement effect on $T_c$, we have performed quantum mechanical computations based on the DFT+U method, where the Hubbard U term is used to treat the strong on-site Coulomb interaction of localized Cu 3*d* electrons, on Bi2201 ($Bi_2Sr_2CuO_{6+\delta}$) and Bi2212 ($Bi_2Sr_2CaCu_2O_{8+\delta}$). The excess oxygen δs are estimated, according to the $T_c$-δ relation, to be 0.17 (27) and 0.25 (28) for Bi2201 ($T_c \sim 10$ K) and Bi2212 ($T_c \sim 90$ K), respectively. Our calculations show that the excess oxygen δ resides energetically favorably between the BiO-layers. Both hydrostatic and non-hydrostatic pressures were applied to shrink the supercell volume, while allowing the shape of the supercell to relax. The experimental compressibility values of the *a*- and *c*-axes were used as the input for the non-hydrostatic case (24, 25). We have calculated the total and partial density of states (DOSs) and the band structures near the Fermi level at selected pressures. Fermi-surface-topology changes are induced by pressures above ~ 37 GPa for Bi2201 and above ~ 42 GPa for Bi2212, respectively. The projected electronic-band Cu $3d_{x^2-y^2}$ variations with pressure are displayed in Figs.4a-b. The accompanying evolutions of the band state occupancy with pressures are shown in Fig.4c. The DOSs vary with pressure in a fashion qualitatively similar to that with the $T_c$s. For these compounds, the calculations also clearly show that pressure inhomogeneity has a rather large effect on the electronic transition, as displayed in Fig. 5, i.e., the pressure inhomogeneity tends to induce the transition at lower pressure. The calculated DOSs under non-hydrostatic pressures more closely reproduce the $T_c$-P behavior observed. This effect has been reported experimentally in HTSs previously (29, 30). Our experimental results and theoretical calculations suggest that it is possible to reach $T_c$s even higher than what we obtained in experiments by continuing to push the applied pressure limit.

Detailed structural studies and transport properties associated with the observed superconducting behaviors are currently being conducted. It would also be interesting to investigate the scaling between $T_c$ and the superfluid density (31) under high pressure in the BSCCO system.

In conclusion, we have observed that the $T_c$s of layered cuprates Bi2201 and Bi2212 increase beyond the maximum $T_c$ predicted by the universal $T_c$-P relation under high pressures, independent of the number of $CuO_2$-layers per formula. We have attributed the $T_c$-resurgence to a pressure-induced electronic transition in the $CuO_2$-layers of the cuprate HTSs, e.g., a Fermi surface topology change, in qualitative agreement with our DFT+U calculations. The observations, therefore, provide a novel path to higher $T_c$s than previously achieved in layered cuprates by breaking away from the universal $T_c$-P relation. They will certainly stimulate future experimental and theoretical investigations on the $T_c$ resurgence in other cuprates, as well as in Fe-based superconductors.

## Materials and Methods

**Crystal Growth.** The polycrystalline Bi2201 was prepared by the standard solid-state method followed by a prolonged oxygen annealing. The stoichiometric mixture of high purity $Bi_2O_3$ (99.9995%), $SrCO_3$ (99.99%), and CuO (99.995%) was pressed into a pellet and then sintered in a temperature range of 700 °C to 850 °C in oxygen atmosphere. This procedure was repeated at least four times with intermediate grinding. It displayed a superconducting transition around 10 K similar to that previously reported (27). The single-crystalline Bi2212 was grown by the floating zone technique (32).

**DC Magnetization Measurements under High Pressure.** We developed a miniature diamond-anvil-cell (mini-DAC) fabricated from BeCu alloy with length ~ 29 mm and outer diameter ~ 8.8 mm, which was adapted into a Quantum Design Magnetic Property Measurement System (MPMS) for ultra-sensitive magnetization measurements at temperatures down to 1.8 K and in fields up to 7 T under high pressure. A pair of 300 μm diameter culet-sized diamond anvils was used for all of the measurements under high pressure. The gaskets are made from nonmagnetic Ni-Cr-Al alloy. Each gasket was pre-indented to ~25 – 35 μm in thickness and a 100 – 120 μm diameter hole was drilled to serve as the sample chamber. Thermal grease was used as the pressure-transmitting medium. The applied pressure was measured by the fluorescence line of ruby powders and the Raman spectrum from the culet of the top diamond anvil. Samples with diagonal ~ 80 – 100 μm and thickness of a few micrometers were prepared to provide a sufficient magnetic signal. The reproducibility of the background signal under different magnetic fields and thermal cycles was confirmed. The variance of different runs of background measurements is below $2\times10^{-8}$ emu/Oe, which offers a good platform to analyze the superconducting signal under high pressures. A piece of calibration Pb sample with diagonal ~ 100 μm was tested with the mini-DAC at ambient pressure. A sharp superconducting transition at 7.2 K was detected.

**Density Functional Calculations.** The Vienna ab initio simulation package (VASP) (33, 34) with the projector-augmented wave (PAW) method (35) was used for all of the calculations in this study. The plane wave cutoff energy was set to 450 eV, and the generalized gradient approximation (GGA) with the semilocal Perdew-Burke-Ernzerhof (PBE) (36) function was adopted to describe the exchange correlation interactions. The Brillouin zone was sampled with Γ-centered k-meshes with a spacing of 0.027 Å$^{-1}$ for the crystal structure optimization and 0.013 Å$^{-1}$ for the electronic structure calculations. The tetrahedron method with Blöchl corrections was carried out to obtain the accurate density of states. The criteria of convergence for energy and force were set to $10^{-5}$ eV and 0.005 eV/Å, respectively. On-site Coulomb repulsion of Cu 3*d* electrons was corrected by the GGA+U method (37) with $U_{eff}$ = 4 eV according to previous theoretical reports (38).

In the hydrostatic model, lattice parameters under different pressures are obtained by shrinking the supercell under free relaxation of the lattice parameters with fixed volumes. Under these conditions, the pressures along each direction are the same, i.e. homogeneous pressure. However, the difference between our theoretical lattice parameters under hydrostatic conditions and the reported lattice parameters measured by synchrotron XRD (24, 25) indicates that the actual pressure along the *ab*-plane is smaller than that along the *c*-axis above a certain pressure, which is affected by the hydrostatic limit of the pressure-transmitting medium as well as the strength anisotropy of the testing material. To better simulate the inhomogeneous pressure condition, we used the experimental data as a reference to adjust the pressure along different directions accordingly.

**Acknowledgements** We thank M. Eremets for advice in developing the mini-DAC, and J. B. Zhang, X. J. Chen, R. Dumas, and S. Li for helpful discussions. The work performed at the Texas Center for Superconductivity at the University of Houston is supported by U.S. Air Force Office of Scientific Research Grant FA9550-15-1-0236, the T. L. L. Temple Foundation, the John J. and Rebecca Moores Endowment, and the State of Texas through the Texas Center for Superconductivity at the University of Houston. The work performed at the Department of Materials Science & Engineering at the University of Texas at Dallas is funded in part by the International Energy Joint R & D Program (No. 20168510011350) of the Korea Institute of Energy Technology Evaluation and Planning (KETEP) grant funded by the Ministry of Knowledge Economy, Korean government.

**Author Contributions** L.Z.D. and C.W.C. designed research; L.Z.D. and Z. W. performed research; Y.P.Z., H.C.W, H.Y.S.Y, and Y.F.N. contributed new reagents/analytic tools; L.Z.D., Y.P.Z, K.J.C., and C.W.C. analyzed data. C.W.C. and L.Z.D. wrote the paper.

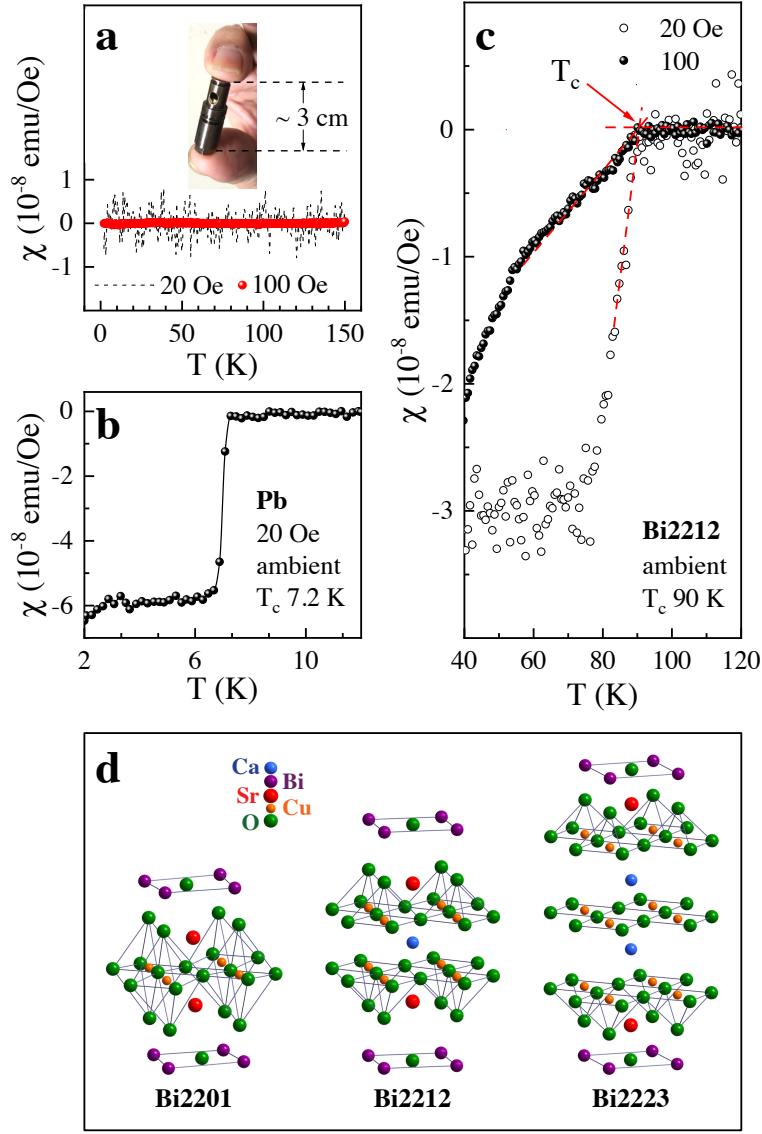

Fig. 1. **DC susceptibility measurements with a miniature diamond-anvil-cell at ambient pressure. a**. The homemade mini-DAC with diameter ~ 8.8 mm and length ~ 3 cm was developed to be adapted to the Quantum Design-MPMS 3 for ultra-sensitive DC magnetization measurements. Background measurements under 20 and 100 Oe before loading samples into the mini-DAC. The variance of different runs of background measurements is below $2\times10^{-8}$ emu/Oe. **b**. Measurement of a calibration Pb sample with diagonal ~ 100 μm and thickness below 10 μm at 20 Oe. A sharp superconducting transition at 7.2 K was detected. **c**. Measurement of a Bi2212 single crystal with diagonal ~ 80 μm and thickness below 10 μm. A superconducting transition at 90 K was observed for measurements under both 20 and 100 Oe, while the 100 Oe measurement shows a better signal-to-noise ratio. We have therefore chosen the onset $T_c$ at 100 Oe as the transition temperature $T_c$ in this study. The red dashed lines are used for $T_{c\text{-onset}}$ extrapolation. **d**. Schematic of the crystal structure of $Bi_2Sr_2Ca_{n-1}Cu_nO_{2n+4+\delta}$ with n = 1, 2, and 3.
9

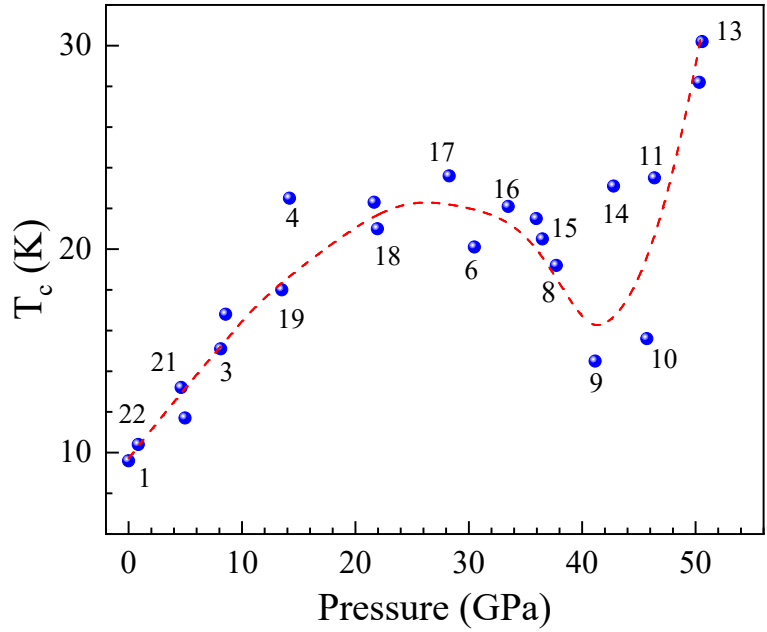

Fig. 2. **Pressure dependence of $T_c$ in under-doped $Bi_2Sr_2CuO_{6+\delta}$**. The $T_c$s were determined by DC magnetization measurements. The numbers represent the sequential order of the experimental runs during both loading and unloading processes. The $T_{c-max}$ is found to be ~ 23 K at a pressure of ~ 26 GPa. $T_c$ at pressures above 40 GPa rises rapidly without any sign of saturation up to 30 K at 51 GPa, which is the highest pressure applied. The red dashed line is a guide for the eyes.

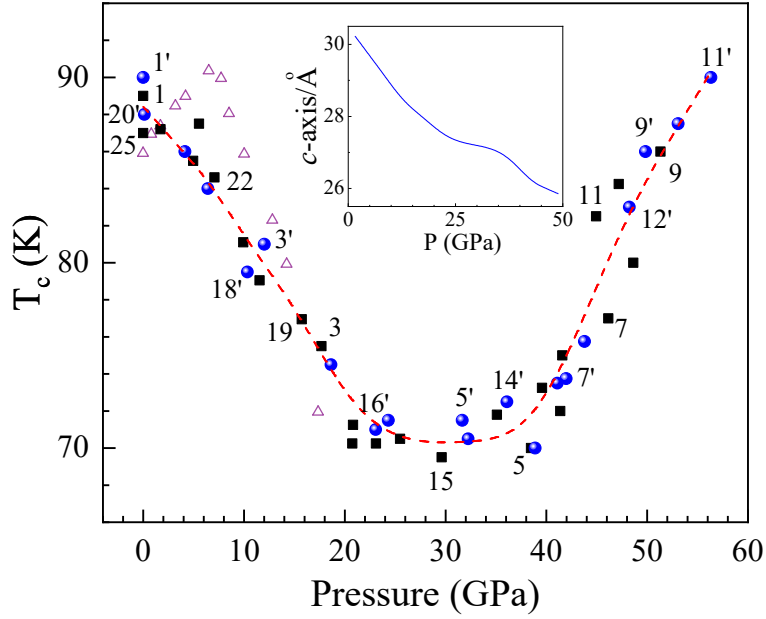

Fig. 3. **Pressure dependence of $T_c$ in slightly over-doped $Bi_2Sr_2CaCu_2O_{8+\delta}$.** $T_c$s were determined by DC magnetization measurements. The two sets of numbers (without and with primes) represent the sequential orders of the experimental runs for two single-crystal samples, with squares representing sample 1 and circles representing sample 2, respectively. The triangles are from Ref. 23 for slightly under-doped Bi2212. After passing a U-shaped valley under pressure of 20-36 GPa, $T_c$ rises rapidly without any sign of saturation to ~ 90 K at ~ 56 GPa, which is the highest pressure applied. The red dashed line is a guide for the eyes. The inset shows the *c*-lattice parameter of Bi2212 under pressures up to 50 GPa from Ref. 24.

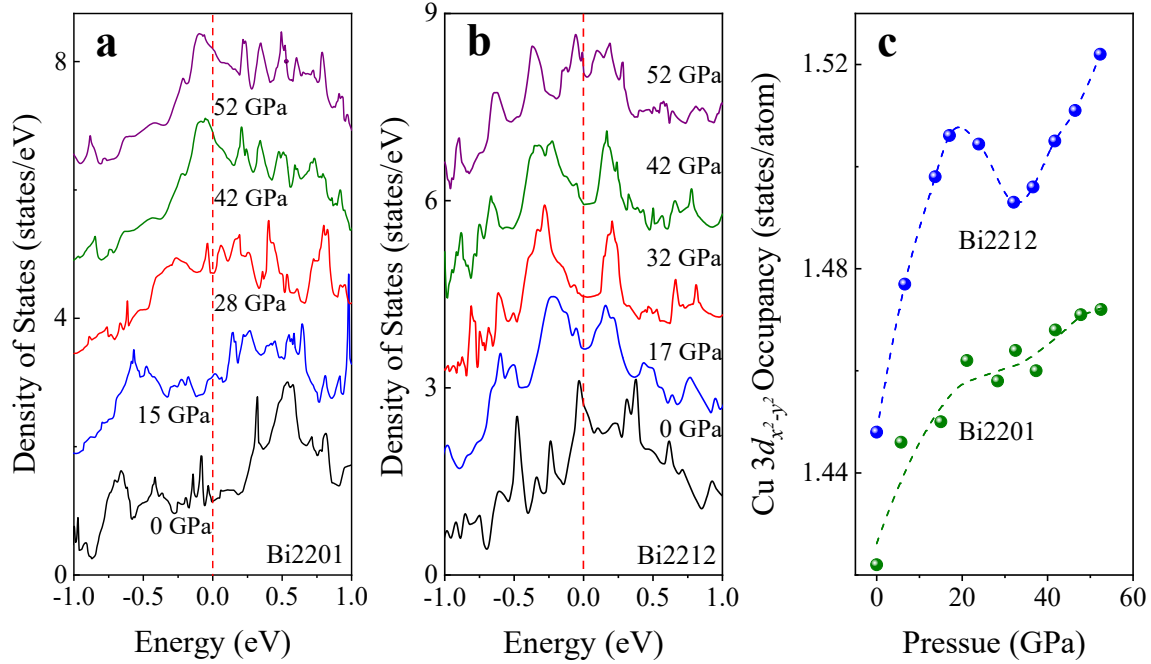

Fig. 4. **Evolution of the projected Cu $3d_{x^2-y^2}$ bands under pressure.** Band shiftings with respect to the Fermi level under pressures from ambient up to 52 GPa: **a**. $Bi_2Sr_2CuO_{6.17}$ (Bi2201) and **b**. $Bi_2Sr_2CaCu_2O_{8.25}$ (Bi2212). Curves are shifted up vertically. **c**. Pressure dependence of band states occupancy, defined as the total states below the Fermi level of the projected Cu $3d_{x^2-y^2}$ band, in Bi2201 and Bi2212. A charge transfer process exists during the compression of the Cu-O bond, which causes the Fermi level right-shifting to allow more occupied states. The density of states (DOSs) vary with pressure in a fashion qualitatively similar to that of the $T_c$s. The Fermi level is set to zero.

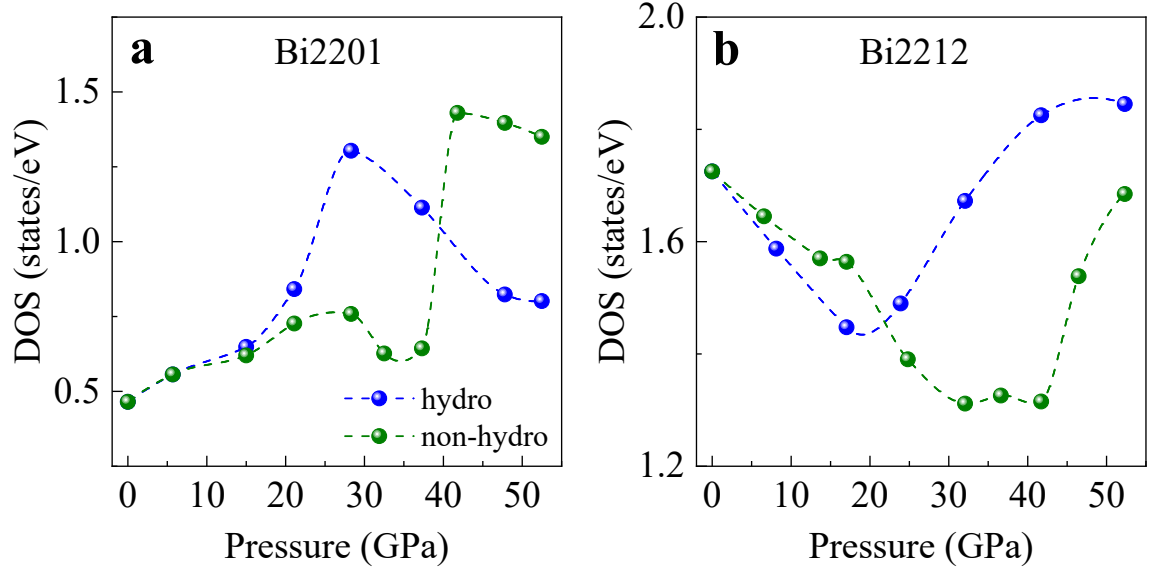

Fig. 5. **Evolution of density of states at the Fermi level under pressure.** Total electronic density of states (DOSs) under hydrostatic (blue) and non-hydrostatic (green) conditions: **a**. $Bi_2Sr_2CuO_{6.17}$ (Bi2201) and **b**. $Bi_2Sr_2CaCu_2O_{8.25}$ (Bi2212). The calculated DOSs under non-hydrostatic pressures more closely reproduce the $T_c$-P behavior observed.